\documentclass[a4paper,10pt]{article}
\usepackage{cite}

\title{ Non-Local Deformation of a Supersymmetric Field Theory}
\author{Qin Zhao$^1$,  Mir Faizal$^{2, 3}$, Mushtaq B Shah$^4$, Anha Bhat$^5$, 
Prince A Ganai$^4$,\\  Zaid Zaz$^6$, Syed Masood$^7$, Jamil Raza$^7$, 
Raja Muhammad Irfan$^7$   \\  \\
$^1$Department of Physics, National University of Singapore, \\
2 Science Drive 3, Singapore 117551\\
$^2$Department of Physics and Astronomy, \\ University of Lethbridge, \\ Lethbridge, AB T1K 3M4, Canada \\
$^3$Irving K. Barber School of Arts and Sciences,
\\ University of British
Columbia - Okanagan \\
  Kelowna,  British Columbia V1V 1V7, Canada
\\$^4$ Department of Physics, \\ National Institute of Technology, \\
Srinagar, Kashmir-190006, India\\ 
$^5$Department of Metallurgical and Materials Engineering, \\ National Institute of Technology, \\ Srinagar 190006, India
\\
$^6$Department of Electronics and Communication Engineering, \\University of Kashmir, \\
Srinagar, Kashmir-190006, India\\ $^7$Department of Physics,
International Islamic University,\\ 
H-10 Sector, Islamabad, Pakistan
}
\date{}
\begin{document}

\maketitle

\begin{abstract}
In this paper, we will analyse a supersymmetric field theory deformed by  generalized uncertainty principle and Lifshitz scaling. It will be observed that this deformed supersymmetric field theory contains  non-local fractional derivative terms. In order to construct such deformed $\mathcal{N} =1$ supersymmetric theory, a harmonic extension of functions will be used. However, the supersymmetry will be only preserved for a free theory and will be broken by the inclusion of interaction terms.  
\end{abstract}
\section{Introduction}
Three dimensional supersymmetry is important as it has been 
 observed in Kondo effect \cite{18a}-\cite{a18}. The original Kondo effect
describes a defect interacting with a free fermi liquid of itinerant 
electrons, and the supersymmetry is introduced if 
the ambient theory is an interacting CFT. In fact,  this introduces qualitatively new features into the system.
 A meta-magnetic transition in models for heavy fermions has been analysed using 
 a doped Kondo lattice model in two dimensions 
 \cite{16}. It has been demonstrated that such a system 
 exhibits a field-driven quantum phase transitions due to a  breakdown of the
Kondo effect \cite{a1}-\cite{a12}. Such systems are analysed using 
 Lifshitz theories  which are theories  based on an anisotropic scaling between space and time. 
 The second order quantum phase transition has also been analysed using Lifshitz theories \cite{1}-\cite{4}. 
The location of a Fermi-surface-changing Lifshitz transition is determined by carrier doping
in some heavy fermion compounds   \cite{15}. 
The chemical potential does not cause a heavy band to shift rigidly due to a strong correlation. This  
is determined by the interplay of heavy and additional light bands crossing the Fermi level.

Three dimensional supersymmetry have also been observed in graphene \cite{s5}-\cite{s4}. Furthermore, 
the van der Waals and Casimir interaction, between graphene 
and a material plate,  between a single-wall carbon nanotube and a plate, between graphene and an atom or a molecule, have been analysed  using Lifshitz scaling  \cite{a5}.  
It may be noted that by generalizing the usual Lifshitz theory, it is possible to  describe such materials which could not be described with the local dielectric response \cite{a3}. The 
Casimir-Lifshitz free energy, between two parallel plates made of dielectric material possessing a constant conductivity at low temperatures,  has been studied; and the temperature correction for this system has also been analysed \cite{a4}. Many properties of narrow heavy fermion bands can be described by a 
Zeeman-driven Lifshitz transition \cite{a2}. The fermionic theories with $ z = 3$ have  been analysed  \cite{2a}-\cite{3a}. 
In fact, the  Nambu-Jona-Lasinio type four-fermion coupling at the $z=3$ Lifshitz  fixed point in four dimensions is asymptotically free and generates a mass scale dynamically \cite{5a}. Furthermore, fermionic theories with  $z=2$ have been constructed, and it has been demonstrated that the  construction of such fermionic theories requires a non-local differential operator \cite{6a}. However, it is possible to analyse this non-local differential operator using the harmonic extension of  functions   \cite{7a}-\cite{12a}. 
 
The Lifshitz theories based on the generalized uncertainty principle  have  also been constructed  \cite{field}. 
The generalized uncertainty principle is motivated by the existence of a minimum length scale, which in turn is predicted from almost all approaches to quantum gravity. According to most quantum gravity theories, the classical picture of spacetime  as a continuous differential manifold  breaks down below the Planck length. This is because fluctuations in the geometry of order one at the Planck scale impose a minimum length scale below which space cannot be probed. Such a minimum measurable length scale 
 occurs in string theory, as space cannot be probed below the string length scale in perturbative string theory 
 \cite{unz2}-\cite{un2z}. In  loop quantum gravity,  the existence of a minimum length 
 turns the big bang into a big bounce \cite{unz1}. Even though the existence of a minimum measurable length scale in predicted in almost
 all theories of quantum gravity, it is not consistent with the usual 
 Heisenberg uncertainty principle. This is because according to the usual  Heisenberg uncertainty principle
  length can in principle be measured with arbitrary precision, if the momentum is not 
 measured  \cite{unzasaqsw,un1,un11, un12, un13, un14, un15, un17,un18,un19,un10,un51,un52,un5}.  So, according to the usual  Heisenberg uncertainty principle, a minimum measurable 
  length scale does not exist.  Therefore, 
 it is necessary   to modify  the Heisenberg uncertainty principle  to make it consistent with the existence of a  minimum measurable length scale.
 This modified   uncertainty principle is called the generalized uncertainty principle. 
 The  modification of the Heisenberg uncertainty principle  leads to a deformation of the usual  Heisenberg algebra.  
 
 Even though the generalized uncertainty principle is motivated from quantum gravity, a modification of this
 principle can have low energy effects 
 which can be detected in laboratory \cite{un54}. In fact, such effects are expected to be observed in 
 Lamb shift, Landau levels, and the tunneling current in a scanning tunneling microscope \cite{un}. 
 Futhermore, as marti have  been recently studied in graphene, it is expected that such a low energy effect from 
 generalized uncertainty principle can be observed in graphene. 
   Thus, it is both interesting and important to analyse supersymmetric theories in three 
 dimensions, with Lifshitz scaling based on the generalized uncertainty principle. 
 Such an analysis would be important to analyse 
 the low energy effect of generalized uncertainty principle on Kondo effect in heavy metals, and van der Waals and Casimir interaction in graphene.  
It will be possible to construct a free  supersymmetric    theory  based on generalized uncertainty principle and  
Lifshitz scaling. Even though  the introduction of interactions will breaks the  supersymmetry of  
 such a theory, such a theory might be interesting as 
 free field theories are also very important as effective field theories to describe materials like graphene. 
 It may be noted that four dimensional supersymmetric theories with Lifshitz scaling have been studied \cite{lifs}-\cite{lifs1}, but 
 so far three dimensional theories with Lifshitz scaling have not been studied. Furthermore, the generalized uncertainty principle has 
 never been combined with supersymmetric field theories based on Lifshitz scaling. However, such a construction is important 
 to analyse condensed matter systems. So, in this paper, we will analysed three dimensional supersymmetric field Lifshitz theories based 
 on the generalized uncertainty principle.

\section{ Deformed  Superspace}
 In this paper, we shall analyse supersymmetric  Lifshitz theories with the existence of a minimum measurable length scale. 
 
 Let us first introduce these two concepts. First, the existence of  minimum measurable length scale is manifested by deforming the usual uncertainty principle to a 
 generalized uncertainty principle, 
\begin{equation}
\Delta x \Delta p = \frac{1}{2} [1 + \beta (\Delta p)^2],
\end{equation}
where   $\beta = \beta_0 \ell_{Pl}^2  $,   $\beta_0$ is a constant normally assumed to be of order one,
and  $\ell_{Pl} \approx 10^{-35}~m$. 
This  deformation of the uncertainty principle in turn deforms  the usual Heisenberg algebra to
\begin{equation}
[x^i, p_j ] =
i   [\delta_{j}^i + \beta p^2 \delta_{j}^i + 2 \beta p^i p_j].
\end{equation}
Correspondingly, the 
   coordinate representation of the momentum operator is modified to the first order in $\beta$ as,  
\begin{equation}
 p_i = -i  \partial_i  (1 -  \beta   \partial^i \partial_i  ).
\end{equation}

Second, in theories with Lifshitz scaling, space and time scale differently. Thus, we can write the scaling of 
 space and time   as 
 \begin{eqnarray}
 x \to bx, \nonumber \\ t \to b^z t,
\end{eqnarray} 
where $z$ is called the degree of anisotropy and $b$ is called the scaling factor. 
In this paper, we shall consider $z =2$. 
It may be noted that this transformation reduces to the usual  conformal transformation for 
$z =1$. 

Now we will incorporate the generalized uncertainty principle into a theory with Lifshitz scaling. Such deformed three dimensional Lifshitz bosonic action is given by \cite{field}
\begin{eqnarray}
S_{b}&=&  \frac{1}{2}\int d^3 x~\left(
\phi \partial^0 \partial _{0}\phi- \kappa ^{2} \partial ^{i}\phi \mathcal{T}^2_{\partial}   
 \partial _{i}\phi \right), 
\end{eqnarray} 
where the non-local fractional derivative operator $\mathcal{T}_{\partial}$ is given by 
\begin{eqnarray}
  \mathcal{T}_{\partial}   &=& T_\partial (1 - \beta \partial^j \partial_j)
 \nonumber \\ &=& \sqrt{-\partial^i \partial_i} (1 - \beta \partial^j \partial_j). 
\end{eqnarray}
Such incorporation breaks the Lifshitz scaling, 
as $\beta$ does not scale with the space and time. However, it is possible to preserve the Lifshitz scaling by 
promoting  the parameter $\beta$ to a background field which scales as \cite{field} 
\begin{equation}
\beta \to b^2 \beta.
\end{equation} 

It may be noted that the non-local differential operator used in the construction of the Lifshitz bosonic action 
based on the generalized uncertainty principle can be analysed using the  harmonic extension of  functions from 
$R^2$ to $R^2 \times (0, \infty)$ \cite{6a}-\cite{12a}.  In fact, it can be effectively viewed as a local 
differential operator by using this harmonic  extension of functions.  
The operator   $ {T}_{\partial}$ can be defined by its action on 
 functions 
 $f: R^2 \to R $. In this case,    its  harmonic extension $u: R^2\times (0, \infty) \to R$ satisfies, 
$
{T}_{\partial} f(x) = -\partial_y u (x, y)| _{y =0}
$. Now let  $u: R^2 \times (0, \infty) \to  R$ be the  harmonic extension 
of $f: R^2 \to R$,  such that
 its restriction 
to $R^2$ coincides with  $f: R^2 \to R$.  Now the solution of the Dirichlet problem defined by  $u(x, 0) = f(x)$ and $ \partial^2  u (x, y) =0  $,
can be used to find $u$, where  
$\partial^2$ is the Laplacian on $R^3$. 
There exists a   unique  harmonic extension 
 $ u \in C^\infty (R^2 \times (0, \infty))$  for a smooth function $C^\infty_0 (R^2) $. 
 Now we can write  $
 {T}_{\partial}^2 f (x) = \partial^2_y u(x, y)|_{y =0}
= - \partial^i \partial_i u(x, y)|_{y =0}$, because  $ {T}_{\partial} f (x)$ also has a harmonic
extension to $R^2 \times (0, \infty)$. 
Furthermore, it is possible to write   $ {T}_{\partial} = \sqrt{- \partial^i \partial_i}$,   
as $ {T}_{\partial}^2 f(x) = - \partial^i \partial_i  f(x)$. 
Thus, we obtain  $ {T}_{\partial} \exp ikx  = |k| \exp ikx$, as 
$ {T}_{\partial}^2 \exp ikx  = |k|^2  \exp ikx$. 

Now using this scalar product, we can write 
 the bosonic action as 
\begin{equation}
S_b =\frac{1}{2}\int d^{3}x~i\partial ^{\mu }\phi ~G_{\mu \nu }
\partial ^{\nu }\phi,
\end{equation} 
where $G_{\mu\nu}$ is  a matrix. 
It is also possible to define a 
 set of local gamma matrices  such that they statisfy 
\begin{equation}
  \{ \Gamma_\mu, \Gamma_\nu\} = 2 G_{\mu\nu}.
\end{equation}
It is possible to write a Lifshitz 
 fermionic  operator based on generalized uncertainty principle as
\begin{equation}
\Gamma^\mu \partial_\mu = 
 \gamma^0 \partial _{0 } +  \gamma^i \kappa \mathcal{T}_{\partial} \partial _{i }.
\end{equation}
This is because if $
  \{ \gamma_\mu, \gamma_\nu\} = 2 \eta_{\mu\nu}
$, then it is possible to write 
$\Gamma_0 = \gamma_0$ and $\Gamma_i = \kappa \mathcal{T}_{\partial}  \gamma_i$. 
Furthermore,  we can also write 
\begin{equation}
 \Gamma^\mu \partial _{\mu }
 \Gamma^\nu \partial _{\nu } = \partial^0\partial_0 - \kappa^2 (\partial^i\partial_i(1 -\beta \partial^k \partial_k)) ^2.
\end{equation}
We can write a Lifshitz fermionic action based on generalized uncertainty principle using three dimensional spinor fields, 
 $\psi_a = \psi^b C_{ba}, $ and $  \psi^a = C^{ab}\psi_b$. Here we have    
$C_{ab}C^{cd} = \delta^c_a \delta^d_b -\delta^c_b \delta^d_a $. 
The square of these spinor fields is given by  $\psi^2 = \psi^a \psi_a/2$. 
Now the Lifshitz fermionic action based on generalized uncertainty principle can be written as 
\begin{eqnarray}
S_f&=& \frac{1}{2}
\int d^{3}x~\psi^a ( \Gamma^\mu \partial _{\mu })^b_a \psi_b \nonumber \\ 
&=&\frac{1}{2}
\int d^{3}x~ \psi^a ( \gamma^0 \partial _{0 } + \gamma^i \kappa \mathcal{T}_{\partial} \partial _{i })^b_a \psi_b. 
\end{eqnarray}

We have the  Lifshitz bosonic and Lifshitz fermionic theories based on the generalized uncertainty principle, 
and so we can  
can  construct  a free supersymmetric theory with  $\mathcal{N} =1$ supersymmetry using these actions.
Thus,  motivated by the definition of generator of ordinary 
$\mathcal{N} =1$ supersymmetry,  we can  write the  
generator of $\mathcal{N} =1$ supersymmetry for a Lifshitz theory based on the generalized uncertainty as 
\begin{equation}
Q_a = \partial_a - ( \gamma^0 \partial _{0 } \theta +  \gamma^i \kappa \mathcal{T}_{\partial} \partial _{i }\theta)_a. 
\end{equation}
Now let $u (x, y) $ be the  harmonic extension of  $f (x) $, and so
$ \partial_i u (x, y)$ will be the harmonic extension of   $\partial_i f(x)$,  
\begin{eqnarray}
 {T}_{\partial} \partial_i f(x) &=&  -
\partial_y \partial_i u(x, y)|_{y =0} \nonumber \\  &=&  - \partial_i u_y (x, y)|_{y =0}. 
\end{eqnarray} 
Furthermore, we have  
$- \partial_i u_y (x, y)|_{y =0} = \partial_i {T}_{\partial}
f(x)$ as  $ {T}_{\partial} f(x) = - u_y (x, 0)$. 
So, the operator   $ {T}_{\partial} 
$ commutes with an ordinary  derivative $\partial_i$,
 \begin{equation}
  {T}_{\partial} \partial_i f(x) = \partial_i {T}_{\partial} f(x).
 \end{equation}
Thus, we can now construct a super-derivative $D_a$ which will commute with the 
  generator of  $\mathcal{N} = 1$ supersymmetry,  
\begin{equation}
D_a = \partial_a - ( \gamma^0 \partial _{0 } \theta -  \gamma^i \kappa \mathcal{T}_{\partial} \partial _{i }\theta)_a. 
\end{equation}
Furthermore, they also obey the following non-local supersymmetric  algebra, 
\begin{eqnarray}
\{Q_a, Q_b\} &=& 
2 ( \gamma^0 \partial _{0 }  + \gamma^i \kappa \mathcal{T}_{\partial} \partial _{i })_{ab},\nonumber \\
\{D_a, D_b\} &=&
- 2 ( \gamma^0 \partial _{0 }  + \gamma^i \kappa \mathcal{T}_{\partial} \partial _{i })_{ab},\nonumber \\
\{Q_a, D_b\} &=&
0.
\end{eqnarray} 
The  states in this theory 
that are invariant under a symmetry are annihilated by 
generators of that symmetry. So, by taking the trace of  $\langle E |\{ Q_a, Q_b \}| E \rangle $, 
it is possible to demonstrate that 
the energy of the 
ground state vanishes even for this deformed supersymmetric theory.  Furthermore, as the 
   Lifshitz    momentum deformed by the generalized uncertainty principle 
   again commutes with the generators of the supersymmetry, there 
occurs a degeneracy in the mass of two states  that are related to 
each other by these generators of supersymmetry.

 However, now because of the non-local differential operator in the definition of 
 $Q_a$, these variations do not obey the Leibniz rule and so the differentiation of 
 a product of superfields is not the same as the differential of each of those superfields. 
 This problem can be evaded for free theories. This is because for free theories we can always 
 shift one differential operator at a time from one field to another in the Lagrangian. Thus, 
 in case of free theories, even theories with  Lifshitz  scaling deformed by generalized uncertainty principle,  we can still construct a non-local supersymmetric  field theory using 
 superspace formalism.
 But as soon as the interactions are introduced, they will tend to break this supersymmetry. 
Now we will analyse some properties of the superspace which is suitable to construct free non-local supersymmetric  theories. 
First, we have 
\begin{equation} 
 D_a D_b = - C_{ab} D^2 -  ( \gamma^0 \partial _{0 }  + \gamma^i \kappa \mathcal{T}_{\partial} \partial _{i })_{ab}.
\end{equation}
  Furthermore, the complete anti-symmetrization of three two-dimensional indices vanishes, 
\begin{equation}
2 D_a D_b D_c = D_a \{ D_b , D_c \} +  D_b \{ D_a , D_c\} +  D_c \{ D_a , D_b\}.
\end{equation}
 So,  we can write, $D^a D_b D_a =0$, and $ D^2 D_a = - D_a D^2 $, where 
$
D^2 D_a  = ( \gamma^0 \partial _{0 } D  + \gamma^i \kappa \mathcal{T}_{\partial} \partial _{i } D)_{a}
$. 
These properties will be used to study various non-local Lifshitz supersymmetric  field theories based on the 
generalized uncertainty principle. 

\section{Supersymmetric Field Theory} 
In this section, we will analyse Lifshitz  supersymmetric  field theories based on generalized uncertainty principle. 
We will write an action for a generalized uncertainty principle  deformed Lifshitz  theory in $\mathcal{N} =1$ superspace 
formalism, so that it has manifest 
$\mathcal{N} =1$ supersymmetry. In order to do that, we first expand  a superfield $\Phi$ as 
$
 \Phi = \phi + \psi^a \theta_a - \theta^2 F  
$. 
Now we can write $
  \phi = [\Phi]_|, \,  \psi_a = [D_a \Phi]_|, 
 \, 
  F  = [D^2 \Phi]_|
$, here $'|'$ means that at the end of calculations we set $\theta_a =0$. 
The non-local supersymmetric transformations  generated by $\epsilon^a Q_a$
can be written as 
\begin{eqnarray}
\epsilon^a Q_a \phi &=& 
  - \epsilon^a  \psi_a,
  \nonumber \\
\epsilon^a Q_a \psi_a &=& 
  - \epsilon^b  [C_{ab} F +  ( \gamma^0 \partial _{0 }  + \gamma^i \kappa \mathcal{T}_{\partial} \partial _{i })_{ab}\phi ],
  \nonumber \\
 \epsilon^a Q_a F &=& 
  - \epsilon^a   ( \gamma^0 \partial _{0 }  + \gamma^i \kappa \mathcal{T}_{\partial} \partial _{i })_{a}^b \psi_b.
\end{eqnarray}

We can write a free action for the deformed supersymmetic theory in $\mathcal{N} =1$ superspace as 
\begin{eqnarray}
 S_{free} [\Phi] &=& \frac{1}{2} \int d^3 x D^2 [\Phi D^2 \Phi ]_| \nonumber \\
 &=& \frac{1}{2} \int d^3 x [ D^2 \Phi D^2 \Phi + D^a \Phi D_a D^2 \Phi + \Phi (D^2)^2 \Phi ]_|
  \nonumber \\ 
  &=& \frac{1}{2} \int d^3 x [F^2  + 
  \phi ( \partial^0\partial_0 - \kappa^2 (\partial^i\partial_i(1 - \beta \partial^j \partial_j)) ^2\phi 
  \nonumber \\ &&
 + 
 \psi^a ( \gamma^0 \partial _{0 } + \gamma^i \kappa \mathcal{T}_{\partial} \partial _{i })^b_a \psi_b
 ] 
 \nonumber \\ &=&
 S_a + S_b + S_f, 
\end{eqnarray}
where  $S_b$ is the deformed bosonic action,   $S_f$  is
the deformed fermionic action,   and $S_a$ is the deformed action for the auxiliary field $F$.

In this action, the supersymmetric variations of the temporal parts cancel out as in the 
ordinary supersymmetric field theories. Furthermore, the non-local supersymmetric  
variation of  a part of the bosonic action generates, 
$ \epsilon ^a \psi _a \kappa^2 (\partial^i\partial_i(1 - \beta \partial^j \partial_j)) ^2\phi $, 
and this term exactly cancels with 
a term generated by the non-local supersymmetric  variation of a part of fermionic action. 
The fermionic action contains a non-local part,
$\epsilon^b (\gamma^j \kappa \mathcal{T}_{\partial} \partial _{j })_b^a\phi. 
(\gamma^j \kappa \mathcal{T}_{\partial} \partial _{j })_a^c \psi_c $. 
This does not directly cancel
out with 
the non-local supersymmetric  variation of the bosonic part. However, 
if we view the non-local operator in terms 
of harmonic extensions of  functions, and
then this term can  be written  as
$\epsilon^b \phi  \kappa^2 (\partial^i\partial_i(1 - \beta \partial^j \partial_j)) ^2\psi_b $. 
Here the derivatives only act on the fermionic part. 
Let $u_1(x, y)$ be the harmonic extension of  $f_1 (x)$ to $ C = R^2  \times (0, \infty)$, and $u_2 (x, y)$ be the harmonic extension of 
$f_2: (x)$ to $ C = R^2 \times (0, \infty)$. Now
 both these  these harmonic extensions vanish 
for $|x| \to \infty $ and $|y| \to \infty $, and we can write \cite{5a01}
\begin{equation}
\int_C u_1(x, y) \partial^2 u_2 (x, y) dx dy - 
\int_C  u_2(x, y) \partial^2 u_1 (x, y) dx dy
= 0. 
\end{equation}
Thus, we  obtain 
 \begin{equation}
 \int_{R^2} \left(u_1(x, y) \partial_y u_2 (x, y)  -   u_2(x, y) 
\partial_y u_1 (x, y) \right)\left. \right|_{y =0} dx 
= 0. 
 \end{equation}
This   can be expressed in terms of $f_1 (x) $ and $f_2 (x)$,  
\begin{equation}
 \int_{R^2}\left(f_1(x) \partial_y
 f_2 (x) -  f_2(x)\partial_y f_1 (x)\right)  dx 
= 0. 
 \end{equation}
Thus,    $\mathcal{T}_{\partial}$ is  moved from $f_2 (x)$ to $f_1 (x)$,  
 \begin{equation}
 \int_{R^2} f_1 (x) \mathcal{T}_{\partial} f_2 (x) =  
 \int_{R^2} f_2 (x) \mathcal{T}_{\partial} f_1 (x).
 \end{equation}
Now the non-local term generated by the non-local supersymmetric 
variation of the  fermionic action can be expressed in terms of 
$ \epsilon ^a \phi \kappa^2 (\partial^i\partial_i(1 - \beta \partial^j \partial_j)) ^2\psi_a $, and so it 
also cancels out with the non-local supersymmetric  variation of the bososnic action. 
It may be noted that this can be done only formally by using the theory of
harmonic extensions of  functions from 
$R^2$ to $R^2 \times (0, \infty)$. Similarly, the remaining terms generated by non-local supersymmetric  
variation of the fermionic part cancel with the terms generated by the non-local supersymmetric  variation 
of the auxiliary field. 
This theory will  have a generalized uncertainty principle  deformed Lifshitz  scaling and $\mathcal{N} =1$ supersymmetry,
even after the following  mass term, $m D^2 [ \Phi^2 ]_|/2= m\psi^2 + m AF$,
is added to its the Lagrangian. It is possible to show that this mass term is also invariant under the non-local supersymmetric  transformations. 
This is because the invariance of the temporal part is again similar 
to the usual non-local supersymmetric  theories and the invariance of the remaining part can 
be demonstrated by using the theory of harmonic extensions of  functions from 
$R^2$ to $R^2 \times (0, \infty)$, as in the previous case. 

We can now use the standard method --- the functional integral to quantize the supersymmetric Lifshitz free 
field theory deformed by the  generalized uncertainty principle. If it was possible to extend to an
interactive  theory, we could also obtained the Feynman graphs using this method. 
However, it will be demonstrated   that  the interactions terms break the 
supersymmetry in those theories. 
The generating functional integral for the free theory can be written as  
\begin{equation}
 Z_0 [J] = \frac{ D\Phi \exp i \left( S_{free}[\Phi]+ 
 J\Phi \right)}{ D\Phi \exp i \left( S_{free} [\Phi]\right)}, 
\end{equation}
where 
\begin{equation}
J\Phi = \int d^3 x D^2 [J \Phi)]_|.
\end{equation}
Thus, we obtain 
\begin{equation}
Z[J] = \exp - i \int d^3 x D^2[ J (D^2 +m)^{-1}J]_|. 
\end{equation} 
Now  the superfield propagator can be written as 
\begin{equation}
 \langle \Phi (p, \theta_1)  \Phi (-p, \theta_2) \rangle = 
 \frac{D^2 - m}{ p^0p_0 - \kappa^2 (p^ip_i (1 -\beta p^k p_k)) ^2 - m^2} 
 \delta( \theta_1- \theta_2).
\end{equation}

 It may be noted that if we add any interaction term will break the supersymmetry of this theory. 
 This is because even though for a free field theory the non-local derivative can be shifted from one field to the 
 another by using harmonic extensions of  functions from 
$R^2$ to $R^2 \times (0, \infty)$, the Leibniz rule does not hold in general. Thus, 
when we have interacting theories, the non-local supersymmetric  variation of a product of more than 
two fields is not equal to the individual non-local supersymmetric  variation of those fields. 
In fact, if we take a simple interaction of the form, 
\begin{equation}
 S[\Phi] = S_{free}[\Phi] + S_{int}[\Phi], 
\end{equation}
where 
\begin{eqnarray}
S_{int}[\Phi]&=& \frac{\lambda}{6} \int d^3 D^2 [\Phi^3]_| 
\nonumber \\ &=&\frac{\lambda}{2} \int d^3 (\phi\psi^a\psi_a + \phi^2 F ), 
\end{eqnarray}
then it is not invariant under the non-local
supersymmetric variation generated by $\epsilon^a Q_a$. This is because in ordinary supersymmetric 
field theories we need to show that $ \epsilon^a \psi^b (\gamma^\mu \partial_\mu)_{ab}  \phi^2 
=   2 \epsilon^a \psi^b \phi (\gamma^\mu \partial_\mu)_{ab} \phi $, however, for   
the non-local part of this deformed  theory, we have   
$\epsilon^a \psi^b (\gamma^i \kappa \mathcal{T}_{\partial} \partial_i)_{ab}  \phi^2 
\neq   2 \epsilon^a \psi^b \phi (\gamma^i \kappa \mathcal{T}_{\partial} \partial_i)_{ab} \phi$.
Thus, 
the non-local supersymmetric  variation of the interaction terms can not cancel out.

\section{Conclusion}
In this paper,  we analysed a 
supersymmetric theory deformed by generalized uncertainty principle  and    Lifshitz  scaling. The  action of this deformed theory contains  non-local fractional 
derivatives. Thus, even the generators of supersymmetry contain non-local fractional derivative terms. 
However, these fractional derivative terms can effectively be treated as a local  operator by using 
 harmonic extensions of  functions from 
$R^2$ to $R^2 \times (0, \infty)$. Furthermore, this non-local operator commutes with the local 
derivatives, and so  we could  construct a super-derivative which commutes 
with the generator of the supersymmetry. This super-derivative was 
used in the  construction of  various non-local supersymmetric  field theories. A free matter theory
deformed by the  generalized uncertainty principle  and  Lifshitz  scaling was constructed such that it 
 was invariant under non-local supersymmetric  transformations. 
It was argued that any free    non-local supersymmetric  theory will be invariant under non-local supersymmetric transformations. 
However, it was    demonstrated that even a simple 
interaction term will break the supersymmetry of this theory.

 The effect of generalized uncertainty principle 
 on AdS/CFT has already been analysed \cite{faiz}. The AdS/CFT correspondence relates the supergravity solutions on AdS to a superconformal field
 theory on its boundary   \cite{13a}-\cite{17a}. It would be be interesting to analyse the AdS/CFT correspondence for Lifshitz theories based on 
 the generalized uncertainty principle. 
The holographic dual to the Lifshitz field theory has also been analysed \cite{10}-\cite{14}.  
In these Lifshitz theories,    the dependence of 
  physical quantities such as  the energy density  on the momentum scale  is evaluated using the 
 renormalization group flow at finite temperature \cite{b14}. 
In fact, gravity with anisotropic scaling  is obtained from 
the holographic renormalization  asymptotically Lifshitz spacetimes \cite{c14}. 
The holographic counter-terms induced near anisotropic infinity take the
form of the action for gravity at a Lifshitz point. It has been observed that the $z=2$ anisotropic Weyl anomaly in 
dual field theories, in three dimensions, can be obtained from the 
 holographic renormalization of Horava-Lifshitz gravity \cite{a14}. 
In fact, Lifshitz theories have also become important because of the development of Horava-Lifshitz  gravity
\cite{5}-\cite{9}. Even though  the addition of higher order curvature terms to the 
 gravitational action makes it  renormalizable, it spoils the unitarity of this theory. 
 However, it is possible to add 
  higher order spatial derivatives without adding any higher order temporal 
derivatives. 
Even though this  break Lorentz symmetry in the  Horava-Lifshitz theory
of gravity,    General Relativity is recovered the infrared limit.  
It may be noted that a system at finite temperature and finite chemical potential
with a Lifshitz black hole in place of a Lifshitz geometry  has been used for analysing 
the fermionic retarded Green's function  with  $z = 2$ 
\cite{17}. In fact, the  
 Hawking radiation for Lifshitz fermions  has also been studied  \cite{18}. It would be interesting to analyse the effect 
 that generalized uncertainty principle can have on such systems.

\section*{Acknowledgement}
We would like to thank  Ali Nassar for pointing out that the
parameters in a conformal field theory
can be promoted to background fields.  We would also like to thank Douglas Smith for useful discussions on Lifshitz supersymmetry. The work of Q.Z. is
supported by NUS Tier 1 FRC Grant R-144-000-316-112.


\begin{thebibliography}{99}

\bibitem{18a}  S.~Harrison, S.~Kachru and G.~Torroba, Class. Quant. Grav.  29, 194005 (2012)
\bibitem{a18}C. Pepin and M. Lavagna, Phys. Rev. B 59, 12180 (1999)
\bibitem{16}M. Bercx and F. F. Assaad, Phys. Rev. B86, 075108 (2012) 
\bibitem{a1}P. Gegenwart, Nature Phys. 4, 186 (2008)
\bibitem{a12}P. Coleman, C. Pepin, Q. Si and R. Ramazashvili, J. Phys. Condens. Matt. 13, R723 (2001)


\bibitem{1}R. M. Hornreich, M. Luban and S. Shtrikman, Phys. Rev. Lett. 35, 1678 (1975)
\bibitem{2}G. Grinstein, Phys. Rev. B23, 4615 (1981)
\bibitem{3}P. M. Chaikin and T. C. Lubensky, Principles of Condensed Matter Physics,
Cambridge University Press, Cambridge, UK (1995)
\bibitem{4}S. Sachdev, Quantum Phase Transitions,  Cambridge University Press, Cambridge, UK (2001)
\bibitem{15}A. Benlagra and M. Vojta, Phys. Rev. B87, 165143 (2013) 

\bibitem{s5}   E.~M.~C.~Abreu, M.~A.~De Andrade, L.~P.~G.~De Assis, J.~A.~Helayel-Neto, A.~L.~M.~A.~Nogueira and R.~C.~Paschoal,
  JHEP   1105, 001 (2011)
\bibitem{s4}  V. K. Oikonomou, Int. J. Geom. Meth. Mod. Phys. 12, 1550114 (2015)

\bibitem{a5} M. Bordag, B. Geyer, G. L. Klimchitskaya and V. M. Mostepanenko, Phys. Rev. B74, 205431 (2006)


\bibitem{a3}V. B. Svetovoy, Phys. Rev. Lett. 101, 163603 (2008)
\bibitem{a4} S. A. Ellingsen, I. Brevik, J. S. Hoye and K. A. Milton, Phys. Rev. E, 78, 021117 (2008)

\bibitem{a2}A. Hackl and  M. Vojta, Phys. Rev. Lett. 106, 137002 (2011)

\bibitem{2a}D. Anselmi and M. Halat, Phys. Rev. D 76, 125011 (2007)
\bibitem{3a}D. Anselmi, Eur. Phys. J. C 65, 523 (2010) 
\bibitem{5a}A.  Dhar, G. Mandal and S. R. Wadia, Phys. Rev. D80, 105018 (2009)


\bibitem{6a}H. Montani and F. A. Schaposnik, Phys. Rev. D86, 065024 (2012)
\bibitem{7a}L. Caffarelli and L. Silvestre,    Comm. Part. Diff. Eqs.  32, 1245 (2007)
\bibitem{8a}R. T. Seeley, Proc. Symp. Pure Math. 10, 288 (1967)
\bibitem{9a}C. Laemmerzahl, J. Math. Phys. 34, 3918 (1993)
\bibitem{10a}J. J. Giambiagi, Nuovo Cim. A 104, 1841 (1991)
\bibitem{12a}C. G. Bollini and J. J. Giambiagi, J. Math. Phys. 34, 610 (1993)


\bibitem{field} M. Faizal and B. Majumder, Annals  Phys. 357, 49 (2015)

\bibitem{unz2}D. Amati, M. Ciafaloni and G. Veneziano, Phys. Lett. B 216, 41 (1989)
\bibitem{unzasaqsw}  A. Kempf, G. Mangano, and R. B. Mann, Phys. Rev. D 52, 1108 (1995) 
\bibitem{uncsdcas}L.  N.  Chang, D.  Minic, N. Okamura, and T.  Takeuchi, Phys.Rev. D65,  125027 (2002)
\bibitem{uncscds}L.  N.  Chang, D.  Minic, N. Okamura, and T.  Takeuchi, Phys. Rev. D65, 125028 (2002) 
\bibitem{un2z}S. Benczik, L. N.  Chang, D.  Minic, N.  Okamura, S.  Rayyan, and T.  Takeuchi,  Phys. Rev. D66, 026003 (2002)
\bibitem{unz1}P.  Dzierzak, J.  Jezierski, P.  Malkiewicz, and W. Piechocki,  Acta Phys. Polon. B41, 717 (2010) 
\bibitem{un1}     D. Amati, M. Ciafaloni, and G. Veneziano, Phys. Lett. B  216, 41  (1989) 
\bibitem{un11}    M. Maggiore, Phys. Lett. B 304, 65 (1993)
\bibitem{un12}    M. Maggiore,  Phys. Rev. D   49, 5182  (1994)
\bibitem{un13}    M. Maggiore, Phys. Lett. B   319,  83 (1993)
\bibitem{un14}    L. J. Garay,  Int. J. Mod. Phys. A   10,  145 (1995)
\bibitem{un15}    F. Scardigli, Phys. Lett. B   452, 39  (1999)
\bibitem{un17}    C. Bambi, F. R. Urban,  Class. Quantum Grav.   25,  095006 (2008)
\bibitem{un18}    K. Nozari,  Phys. Lett. B.   629,  41 (2005) 
\bibitem{un19}    K. Nozari, T. Azizi,  Gen. Relativ. Gravit.   38, 735  (2006)
\bibitem{un10}    P. Pedram,  Int. J. Mod. Phys. D   19,  2003 (2010)
\bibitem{un51}    A. Kempf, J. Phys. A   30, 2093 (1997) 
\bibitem{un52}    F. Brau,  J. Phys. A   32, 7691 (1999) 
\bibitem{un5}    K. Nozari, and B. Fazlpour, Chaos, Solitons and Fractals,   34, 224 (2007) 
\bibitem{un54}    S. Das, and E. C. Vagenas,  Phys. Rev. Lett.   101,  221301 (2008)
 \bibitem{un}A. F. Ali, S. Das and E. C. Vagenas, Phys. Rev. D   84 , 044013 (2011)
\bibitem{lifs} D. Redigolo,  Phys. Rev. D   85, 085009 (2012) 
\bibitem{lifs1}  S. Chapman, Y. Oz and A. R. Moshe,  JHEP   1510, 162 (2015) 
 \bibitem{5a01} J. Tan, The Brezis–Nirenberg type problem involving the square root of the Laplacian Calc.\ Var\ {\bf 42},\ 21\ (2011)




\bibitem{faiz} M. Faizal. A. F. Ali and A.  Nassar,  Int. J. Mod. Phys. A 30, 1550183 (2015)


 \bibitem{13a}J. M. Maldacena, Adv. Theor. Math. Phys. 2, 231 (1998) 
\bibitem{14a}O. Aharony, S. S. Gubser, J. M. Maldacena, H. Ooguri and Y. Oz,  Phys. Rept. 323, 183 (2000)
\bibitem{15a} E. Witten, Adv. Theor. Math. Phys. 2, 253 (1998) 
\bibitem{16a}J. M. Maldacena and C. Nunez, Phys. Rev. Lett. 86, 588 (2001) 
\bibitem{17a}A. Karch and E. Katz, JHEP.  0206, 043 (2002)
\bibitem{10}S. Kachru, X. Liu and M. Mulligan, Phys. Rev. D78,  106005 (2008)
\bibitem{12}K. Balasubramanian and K. Narayan,  JHEP.  1008, 014 (2010) 
\bibitem{13} R. Gregory, S. L. Parameswaran, G. Tasinato and I. Zavala, JHEP. 1012, 047 (2010) 
\bibitem{14}A. Donos and J. P. Gauntlett,  JHEP. 1012, 002 (2010) 
\bibitem{b14} M. Park and R. B. Mann,  JHEP.  1207, 173 (2012) 
\bibitem{c14}T. Griffin, P. Horava and C. M. M. Thompson, JHEP. 1205, 010 (2012) 
\bibitem{a14}T. Griffin, P. Horava and C. M. M. Thompson, Phys. Rev. Lett. 110,  081602  (2013)

\bibitem{5}P. Horava, Phys. Lett. B 694,172 (2010)
\bibitem{6}P. Horava, Phys. Rev. D79, 084008 (2009)
\bibitem{7}P. Horava, JHEP. 03, 020 (2009)
\bibitem{8}O. Obregon and J. A. Preciado, Phys. Rev. D86, 063502 (2012) 
\bibitem{9}A. Sheykhi, Phys. Rev. D 87, 024022 (2013) 



\bibitem{17}  M.  Alishahiha, M. R.  M.  Mozaffar and A.  Mollabashi,  Phys. Rev. D86, 026002 (2012) 
\bibitem{18} M.  Liu, J. Lu and J. Lu,   Class. Quant. Grav. 28, 125024 (2011) 
 
\end{thebibliography}
\end{document}